%Paper: hep-th/9311002
%From: unel%TRBOUN.BITNET@FRMOP11.CNUSC.FR
%Date: Mon, 01 Nov 1993 12:52:57 +0200

\magnification=\magstep 1
\baselineskip=20pt
\hsize=155truemm
\vsize=235truemm
\hoffset=30truemm
\voffset=30truemm
%\nopagenumbers
%\def\Re{{\rm I}\kern -0.2em{\rm Re}}
%\def\Im{{\rm I}\kern -0.2em{\rm Im}}
%\def\Diff{{\rm I}\kern -0.2em{\rm D}}
%\headline={\hfil\folio}
\vskip 3truecm
\centerline{\bf Unification of gauge couplings in Kaluza-Klein theory}
\centerline{\bf with two internal manifolds}
\vskip 3truecm
\centerline{M. Ar\i k  and V. Gabay}
\vskip 1truecm
\centerline{Bo\u gazi\c ci University,}
\centerline{Centre for Turkish-Balkan Physics Research and Applications,}
\centerline{Bebek, Istanbul, Turkey}
\vskip 3truecm
\centerline{Abstract}
\vskip 1truecm

We consider a Kaluza-Klein theory whose ground state is ${\bf R}^4 \times {\bf
M
} \times {\bf K}$
where ${\bf M}$ and ${\bf K}$ are compact, irreducible, homogenous internal
mani
folds.
This is the simplest ground state compatible with the existence of the
graviton,
gauge fields, massless scalar fields and the absence of the cosmological
constan
t.
The requirement for these conditions to be satisfied are the odd dimensionality
of ${\bf M}$ and ${\bf K}$, and the choice of a dimensionally continued Euler
fo
rm
action whose dimension is the same as the dimension of ${\bf M} \times {\bf
K}$.
 We show
that in such a theory, which is not simple due to presence of two internal
manifolds, the gauge couplings $g^2_M$ and $g^2_K$ are actually unified
provided
that the internal space sizes are constant. For ${\bf M} \times {\bf K} =
S^{2m+
1}
 \times S^{2k+1}$ this gauge coupling unification relation reads $g^2_M / g^2_K
= m / k$.

\vfil\eject

It is generally believed that the ultimate gauge theory of the universe should
b
e
based on a single semisimple gauge group such that it yields a single gauge
coupling for all gauge interactions.$^1$ On the other hand unification of gauge
theories with gravity requires a Klauza-Klein type theory which basically is
a pure gravity theory in high dimensions.$^2$ In Kaluza-Klein theories the
gauge
symmetry arises as a result of isometries of the internal space. Thus the
breaking of the gauge symmetry in Kaluza-Klein theory requires a topological
change of the internal space.$^3$ In this paper we will show that in the
framewo
rk
of Kaluza-Klein theory with two internal spaces, the unification of gauge
couplings can be achieved without unification of the gauge group. The interest
in such theories arises from the fact that the inclusion of massless scalar
fields in nonabelian Kaluza-Klein theory requires the existence of at least
two irreducible internal manifolds. In fact the physical requirements of
presence of the graviton, gauge fields and massless scalar fields and the
absence of the cosmological term in the effective four dimensional theory
can be satisfied provided one chooses a Kaluza-Klein theory whose ground
state is ${\bf R}^4 \times {\bf M} \times {\bf K}$ where ${\bf M}$ and ${\bf
K}$
 are odd
dimensional, compact, irreducible, homogenous manifolds and one uses a
dimensionally continued Euler form action whose dimension is the sum of
the dimensions of ${\bf M}$ and ${\bf K}$.$^4$ In this paper we will show that
t
he
phenomenological conditions of constant internal space size, which is required
if the gauge couplings are time independent and the vanishing of the pressure
in internal space$^5$ which is required if the internal space components of the
energy-momentum tensor are zero yield a unification relation between the
gauge couplings of the isometries of {\bf M} and of {\bf K}.
We start with an action in $D = 4 + (2m+1) + (2k+1)$ dimensions given by
the $2N$-dimensional, dimensionally continued Euler form

$$L^{(N)} = 2^{-N} \ {\delta^{B_1 B_2...B_{2N}}_{A_1A_2...A_{2N}}} \
{R^{A_1A_2}_{B_1B_2}}...{R^{A_{2N-1}A_{2N}}_{B_{2N-1}B_{2N}}}
  \  \eqno{(1)}$$
where $R^{AB}_{CD}$ is the curvature tensor.$^6$ The equations of motion
obtaine
d
from (1), and supplied with an energy-momentum tensor on the right hand side
representing the matter fields read

$$L^{(N) A}_{ \ \ \ B} = - 2^{-N-1} \ {\delta^{B B_1...B_{2N}}_{A
A_1...A_{2N}}}
 \
{R^{A_1A_2}_{B_1B_2}}...{R^{A_{2N-1}A_{2N}}_{B_{2N-1}B_{2N}}} = {T^A_B}
\eqno{(2
)}$$
where $L^{(N) A}_{ \ \ \ B}$ is the $2N$-dimensional Lovelock tensor. For $N =
m
+k+1$ the
reduction of the equations of motion according to a ground state ${\bf R}^4
\tim
es
{\bf M} \times {\bf K}$ where ${\bf R}^4$ is the space-time and ${\bf M}$ and
${
\bf K}$
are internal manifolds yields

$$L^{(m+k+1)}_{\mu\nu} = {(m+k+1)!\over {m! \ k!}} \ E^{(1)}_{\mu\nu} \ H^{(m)}
 \ J^{(k)} = T_{\mu\nu} \ \eqno{(3)}$$

$$L^{(m+k+1)}_{ab} = - \ {(m+k+1)!\over {m! \  k!}} \ \Bigl[ \ E^{(1)} \
H^{(m)}
_{ab} \ J^{(k)} + {m\over 2} \ E^{(2)} \ H^{(m-1)}_{ab} \ J^{(k)}$$
$$+ {k\over2} \ E^{(2)} \ H^{(m)}_{ab} \ J^{(k-1)} \ \Bigr] = 0  \
\eqno{(4)}$$

$$L^{(m+k+1)}_{ij} = - \ {(m+k+1)!\over {m! \  k!}} \ \Bigl[ \ E^{(1)} \
H^{(m)}
 \ J^{(k)}_{ij} + {m\over 2} \ E^{(2)} \ H^{(m-1)} \ J^{(k)}_{ij}$$
 $$+ {k\over2} \ E^{(2)} \ H^{(m)} \ J^{(k-1)}_{ij} \ \Bigr] = 0  \
\eqno{(5)}$
$
where indices in space-time are denoted by $\mu$ , $\nu$ , indices in
internal space ${\bf M}$ are denoted by $a \ , \ b$ and indices in internal
spac
e
${\bf K}$ are denoted by $i \ , \ j$. The Euler forms of space-time, the
interna
l
manifold ${\bf M}$ and internal manifold ${\bf K}$ are denoted respectively
by $E^{(n)}$, $H^{(n)}$ and $J^{(n)}$. Their Lovelock tensors are denoted by
$E_{\mu \nu}$, $H_{a b}$, $J_{i j}$ where $2n$ is equal to the
dimension of the dimensionally continued Euler form and Lovelock tensor.
Any $2N$ dimensional Euler form and Lovelock tensor in $D$ dimensions satisfy

$$L^{(N) A}_{ \ \ \ B} = \bigl({N\over D}-{1\over 2}\bigr) \ L^{(N)} \
\delta^A_B \ \eqno{(6)}$$
provided that the $D$-dimensional space is isotropy irreducible. Using this
expression for ${\bf M}$ and ${\bf K}$, (4) and (5) become

$$2H^{(m)} \ J^{(k)} \ E^{(1)} + \bigl(3mH^{(m-1)} \ J^{(k)} + kH^{(m)} \
J^{(k-1)} \bigr) \ E^{(2)} = 0 $$
$$\eqno{(7)}$$
$$2H^{(m)} \ J^{(k)} \ E^{(1)} + \bigl(mH^{(m-1)} \ J^{(k)} + 3kH^{(m)} \
J^{(k-1)} \bigr) \ E^{(2)} = 0 $$
For isotropy irreducible internal manifolds ${\bf M}$ and ${\bf K}$ the Euler
fo
rms
$H^{(n)} \ (n = m, \ m-1)$ and $J^{(n)} \  (n = k, \ k-1)$ are constant and
scal
e as the
$2n$'th inverse power of the size (radius) of the manifold. $E^{(1)}$ is the
Ricci scalar (two dimensional Euler form) and $E^{(2)}$ is the Gauss-Bonnet
scalar (four dimensional Euler form) of space-time. The compatibility of the
two equations (7) yields

$$ m \ {H^{(m-1)}\over H^{(m)}} = k \ {J^{(k-1)}\over J^{(k)}} \ \eqno{(8)}$$

Because of the scaling property of the Euler forms $H^{(n)} \simeq b^{-2n}$
and $J^{(n)} \simeq c^{-2n}$ where $b$ and $c$ denote the sizes (radii) of
internal spaces {\bf M} and {\bf K} . (8) reads

$$ m \ b^2 \simeq n \ c^2 \ \eqno{(9)}$$
It can be shown that this relation of proportionality, for the case of
spherical internal manifolds becomes an exact equality.
\
In Kaluza-Klein theory the gauge couplings are inversely proportional
to the square of the internal space size provided that one fixes the four
dimensional Newton gravitational constant $G$ . The exact factors to change
these proportionalities into equalities depend on the specific nature of the
internal space. We will write all following equations in such a manner
that proportionality holds for all kinds of isotropy irreducible internal
spaces, whereas equality holds for ${\bf M}$ = $S^{2m+1}$ and ${\bf K}$ =
$S^{2k+1}$.
We write the action as

$$ S = {8\pi\over G' } \ \int \ L_N \ dV \ \eqno{(10)}$$
where $dV$ is the volume element in  $D = 4 + (2m+1) + (2k+1)$  dimensions and
fix the $D$-dimensional gravitational coupling $G' $ such that when this action
is reduced to four dimensions keeping only the graviton, it becomes

$$ S = {8\pi\over G} \ \int \ R \ \sqrt {-det g} \ d^4x \ \eqno{(11)}$$
where $R = E^{(1)}$ is the Ricci scalar of space-time. Comparing (10) and (11)
gives

$$ G' \simeq \ G \ b \ c \ {4\pi^{m+1}\pi^{k+1}\over m! \ k!} \ \ . \
\eqno{(12)
}$$
When a similar reduction is made keeping the gauge boson one obtains

$$ g^2_M \ \simeq \ {G\over {8\pi b^2}} \ \ \ \ and \ \ \ \ g^2_K \ \simeq \
{G\over {8\pi c^2}} \ \ . \ \eqno{(13)}$$
Using (9), (13) becomes

$$ {g^2_M\over{g^2_K}} \ \simeq \ {m\over k} \ \eqno{(14)}$$
where dim ${\bf M}$ = $2m+1$ and dim ${\bf K}$ = $2k+1$ .

This coupling constant
unification relation follows from conventional Kaluza-Klein theory with two
irreducible internal manifolds provided the dimensions of the manifolds
and the action are chosen to satisfy certain phenomenological truths as
mentioned in our introductory paragraph. Other fundamental assumptions which
lead to (14) are that the internal space size is constant and the internal
space components of the energy-momentum tensor vanish. Of course, a theory
which would yield these assumptions instead of needing them is certainly
more desirable. We hope that the considerations in this paper will be useful
for building such a theory.

\vfil\eject

\centerline {References}
\vskip 2cm
\noindent
[1] S.L. Glashow, Rev. Mod. Phys. {\bf 52}, 539 (1980) ; A. Salam,
Rev. Mod. Phys. {\bf 52}, 525 (1980) ; S. Weinberg, Rev. Mod. Phys. {\bf 52},
515 (1980).
\vskip 0.5cm
\noindent
[2] M.J. Duff, B.E.W. Nilsson and C.N. Pope, Phys. Rep. {\bf 130}, 1 (1986) and
references therein.
\vskip 0.5cm
\noindent
[3] M. Ar\i k and M. Mungan, Mod. Phys. Lett. {\bf A5}, 2593 (1990).
\vskip 0.5cm
\noindent
[4] M. Ar\i k and V. Gabay, "Massless scalar fields in non-Abelian Kaluza-Klein
\break theory", Bo\u gazi\c ci University Preprint BUFB92-2, to be published in
\break Int. J. Mod. Phys. A.
\vskip 0.5cm
\noindent
[5] F. M\"uller-Hoissen, Class. Quant. Grav. {\bf 3}, 665 (1986).
\vskip 0.5cm
\noindent
[6] D. Lovelock, J. Math. Phys. {\bf 12}, 498 (1971).

\bye